\documentclass[letterpaper]{article} 
\usepackage{aaai2026}  
\usepackage{times}  
\usepackage{helvet}  
\usepackage{courier}  
\usepackage[hyphens]{url}  
\usepackage{graphicx} 
\def\UrlFont{\rm}  
\usepackage{natbib}  
\usepackage{caption} 
\usepackage[table]{xcolor}
\definecolor{newcolor}{RGB}{221, 221, 238}
\usepackage{multirow}
\usepackage{dsfont}
\usepackage{amsmath}   
\usepackage{amssymb}   
\usepackage{verbatim}
\usepackage{xcolor}
\usepackage{subfig}
\usepackage{algorithm}
\usepackage{algorithmic}
\usepackage{color,xcolor}
\usepackage{amsmath}  

\definecolor{commentcolor}{RGB}{63,127,127}   
\newcommand{\PyComment}[1]{\textcolor{commentcolor}{\texttt{\# #1}}}

\def\ie{\textit{i.e.,}~}
\def\eg{\textit{e.g.,}~}
\newcommand{\figref}[1]{Fig.~\ref{#1}}
\newcommand{\tabref}[1]{Table~\ref{#1}}

\usepackage{newfloat}
\usepackage{listings}
\DeclareCaptionStyle{ruled}{labelfont=normalfont,labelsep=colon,strut=off} 
\floatstyle{ruled}
\newfloat{listing}{tb}{lst}{}
\floatname{listing}{Listing}
\title{SPP-SCL: Semi-Push-Pull Supervised Contrastive Learning for Image-Text Sentiment Analysis and Beyond}

\author{
    Jiesheng Wu\textsuperscript{\rm 1},
    Shengrong Li\textsuperscript{\rm 2}
    \thanks{Corresponding author.}
}

\affiliations{
    \textsuperscript{\rm 1}School of Computer and Information, Anhui Normal University, Wuhu, 241003, China\\
    \textsuperscript{\rm 2}College of Artificial Intelligence, Nanjing University of Aeronautics and Astronautics, Nanjing, 211106, China\\
    jasonwu@mail.nankai.edu.cn, lisrong@nuaa.edu.cn
}

\begin{document}

\maketitle

\begin{abstract}
Existing Image-Text Sentiment Analysis (ITSA) methods may suffer from inconsistent intra-modal and inter-modal sentiment relationships. Therefore, we develop a method that balances before fusing to solve the issue of vision-language imbalance intra-modal and inter-modal sentiment relationships; that is, a Semi-Push-Pull Supervised Contrastive Learning (SPP-SCL) method is proposed. Specifically, the method is implemented using a novel two-step strategy, namely first using the proposed intra-modal supervised contrastive learning to pull the relationships between the intra-modal and then performing a well-designed conditional execution statement. If the statement result is false, our method will perform the second step, which is inter-modal supervised contrastive learning to push away the relationships between inter-modal. The two-step strategy will balance the intra-modal and inter-modal relationships to achieve the purpose of relationship consistency and finally perform cross-modal feature fusion for sentiment analysis and detection. Experimental studies on three public image-text sentiment and sarcasm detection datasets demonstrate that SPP-SCL significantly outperforms state-of-the-art methods by a large margin and is more discriminative in sentiment.
\end{abstract}

\section{Introduction}
\label{S0}

With the explosive growth of social media, users increasingly express sentiments through images accompanied by textual descriptions. This has spurred extensive interest in \textbf{image-text sentiment analysis (ITSA)} \cite{das2023multimodal,xue2022multi,zadeh2017tensor}. The core challenge of ITSA lies in accurately modeling and integrating emotional cues from different modalities to make consistent and robust predictions.

Early methods typically concatenate high-level features from each modality for fusion \cite{xu2017multisentinet,xu2017analyzing}, but such strategies often fall short in capturing nuanced or conflicting emotional signals across modalities. To address this, recent studies have introduced interaction-based fusion modules \cite{yang2020image, yuadapting, 9920172} to better model fine-grained cross-modal sentiment relationships. Despite progress, a critical issue remains underexplored: \textit{the imbalance and inconsistency among intra-modal and inter-modal sentiment representations}.
\begin{figure}[H]
    \centering
    \includegraphics[scale=0.54]{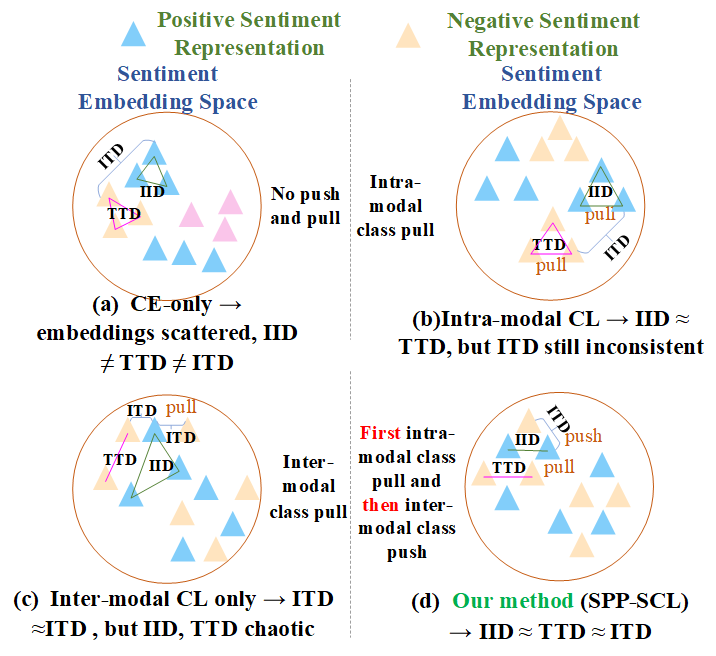}
    \caption{Comparison of intra-modal and inter-modal sentiment distances under different training strategies.
(a) Cross-entropy loss only: image and text sentiment representations are scattered, leading to inconsistent intra- and inter-modal sentiment relationships.
(b) Intra-modal contrastive learning aligns IID and TTD, but fails on ITD.
(c) Inter-modal contrastive learning aligns ITD but ignores intra-modal structure.
(d) Our proposed SPP-SCL balances all three sentiment distances, yielding consistent sentiment embeddings.}
    \label{5-1-1}
\end{figure}
As illustrated in \figref{5-1-1}, existing methods primarily rely on cross-entropy loss, which encourages discriminative classification but fails to align sentiment representations across modalities. Specifically, we observe that the average pairwise distances within image sentiment features (IID), within text sentiment features (TTD), and between image-text pairs sentiment features (ITD) often differ significantly, even for samples sharing the same sentiment label. \textbf{Here, IID (Image-to-Image sentiment embedding Distance), TTD (Text-to-Text sentiment embedding Distance), and ITD (Image-to-Text sentiment embedding Distance) denote the average Euclidean distances between sentiment embeddings of samples with the same label, computed respectively within the image modality, within the text modality, and across modalities.} Such inconsistencies in sentiment-level embedding space undermine generalization and hinder effective fusion. While some works employ contrastive learning (CL) to address this issue \cite{li2022clmlf}, most adopt a \textit{fuse-then-contrast} strategy, applying CL on fused multi-modal features. This approach risks semantic noise propagation and lacks targeted alignment of unimodal emotional representations.

To tackle these challenges, we propose a novel \textbf{Balance-before-Fuse} framework that aligns sentiment representations within and across modalities before fusion. Inspired by alignment-then-fusion paradigms in vision-language tasks \cite{li2021align, li2022blip, yang2021taco}, we introduce a two-step \textbf{Semi-Push-Pull Supervised Contrastive Learning} (SPP-SCL) approach. In the first step, we apply supervised contrastive losses to image and text branches independently, pulling same-sentiment samples closer within each modality and achieving intra-modal sentiment alignment (\ie, $\textrm{IID} \approx \textrm{IID}$ and $\textrm{TTD} \approx \textrm{TTD}$). In the second step, we conditionally apply inter-modal contrastive learning to align sentiment representations across modalities, based on a diagonal similarity threshold. \textbf{We aim to simultaneously minimize the discrepancy among intra-image, intra-text, and image-text \emph{sentiment} distances under a unified supervised contrastive learning framework}, thereby constructing a modality-balanced and sentiment-consistent embedding space that facilitates robust multi-modal fusion. To further improve fusion quality, we introduce two lightweight yet effective components: a \textbf{Hierarchical Attention (HA)} module for context-aware sentiment extraction from text, and a \textbf{Cross-Modal Fusion (CMF)} module for dynamic integration of aligned features. \textit{Unlike general-purpose vision-language alignment frameworks, our method is specifically tailored for sentiment analysis by aligning emotion-aware representations across modalities, rather than generic semantic features.}

\noindent \textbf{Our main contributions are summarized as follows:}
\begin{itemize}
    \item We propose a novel two-step \textbf{Semi-Push-Pull Supervised Contrastive Learning (SPP-SCL)} framework that aligns intra- and inter-modal \emph{sentiment} relationships prior to fusion, yielding a well-structured space.

    \item We design three supervised contrastive objectives—two intra-modal (for image and text) and one inter-modal—that enable fine-grained sentiment-level alignment without requiring data augmentation or large-scale pretraining.

    \item We develop a Hierarchical Attention module for efficient sentiment representation from text, and a Cross-Modal Fusion module for robust visual-linguistic interaction.

    \item Extensive experiments on three public datasets (MVSA-S, MVSA-M, and HFM) demonstrate that our method significantly outperforms state-of-the-art baselines, especially in fine-grained or ambiguous sentiment scenarios.
\end{itemize}

\section{Related Work}
\label{S1}
\subsection{Image-Text Sentiment Analysis}
Image-Text Sentiment Analysis (ITSA) aims to classify the sentiment of paired visual and textual content, particularly in social media contexts \cite{10468627,10124248,9932611}. Early approaches such as HSAN and MultiSentiNet \cite{xu2017analyzing,xu2017multisentinet} extracted features independently from each modality and fused them via simple concatenation, which limited cross-modal interaction. To address this, subsequent works introduced interaction-aware architectures. For example, Co-MN-Hop6 \cite{xu2018co} and MVAN \cite{yang2020image} leveraged memory-based networks to enable iterative alignment between modalities, while MGNNS \cite{yang2021multimodal} used sentiment-aware graph neural networks. More recent methods such as ITIN and MULSER \cite{9736584,9920172} focused on deep fusion but often overlooked token-level alignment or sentiment consistency. To improve alignment, CLMLF \cite{li2022clmlf} adopted contrastive learning using both label- and data-based signals. However, it applied contrastive objectives directly on fused representations, without explicitly balancing intra- and inter-modal features. Inspired by this, our method introduces a \textit{balance-before-fuse} strategy to align intra- and inter-modal sentiment embeddings prior to fusion.

\subsection{Contrastive Learning}
Contrastive Learning (CL) has shown strong representation power in vision and language domains through instance discrimination techniques, as demonstrated in MoCo \cite{he2020momentum}, SimCLR \cite{chen2020simple}, and SimCSE \cite{gao2021simcse}. In the multi-modal setting, CLIP \cite{radford2021learning} paved the way for joint vision-language representation learning. Supervised contrastive learning (SCL) \cite{khosla2020supervised} extends this idea by incorporating label information, enabling multi-positive alignment. Recent applications of SCL include text classification \cite{gunel2020supervised} and trimodal sentiment analysis \cite{mai2022hybrid,10096777}. However, most methods treat contrastive learning as an auxiliary module or apply it after feature fusion, which may propagate semantic noise across modalities. In contrast, we propose a two-step supervised contrastive framework that explicitly aligns intra-modal and inter-modal sentiment embeddings, ensuring a consistent and discriminative feature space for downstream fusion.

\section{Methodology}
\label{S2}
\subsection{Overview}
\begin{figure*}[h]
    \centering
\includegraphics[scale=0.55]{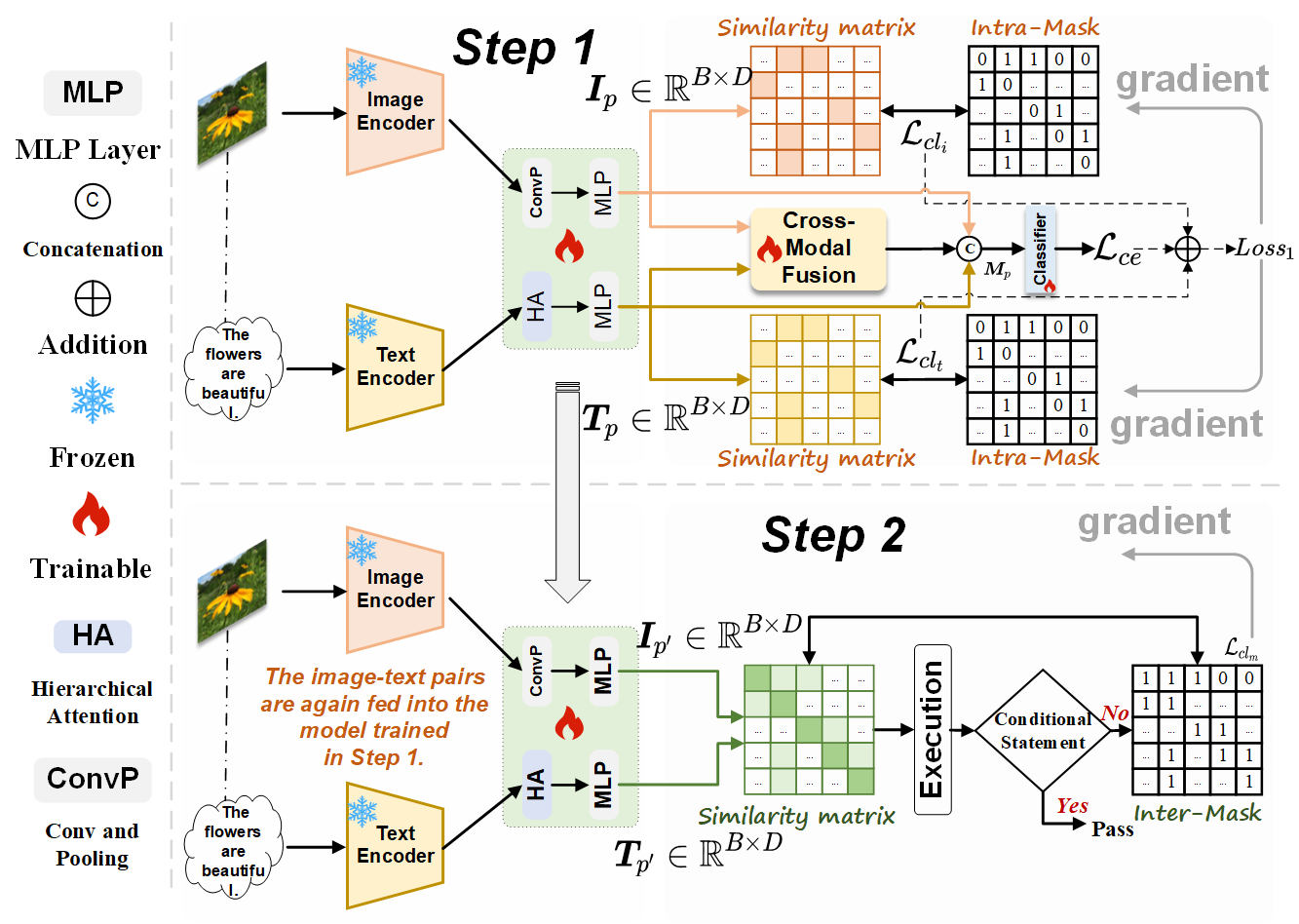}
    \caption{Overall architecture of SPP-SCL. The framework includes two main steps: intra-modal sentiment alignment via supervised contrastive learning ($\mathcal{L}_{cl_{i}}$ and $\mathcal{L}_{cl_{t}}$), and conditional inter-modal sentiment alignment ($\mathcal{L}_{cl_{m}}$).}
    \label{5-2}
\end{figure*}
The overall architecture of our proposed \textbf{Semi-Push-Pull Supervised Contrastive Learning} (SPP-SCL) framework is illustrated in \figref{5-2}. The model operates in two sequential steps designed to align sentiment representations both within and across modalities, prior to final fusion. We first adopt two frozen feature encoders—a ResNet-50 pretrained on ImageNet for image inputs and a BERT-base model pretrained on general text corpora for textual inputs—to extract high-level semantic features. These encoders are \textbf{kept fixed during training} to prevent overfitting on small-scale sentiment datasets and to ensure that performance improvements stem primarily from our proposed contrastive alignment strategy rather than feature extractor fine-tuning. In the \textbf{first step}, intra-modal sentiment alignment is performed using two supervised contrastive losses: $\mathcal{L}_{cl_{i}}$ for the image branch and $\mathcal{L}_{cl_{t}}$ for the text branch. These losses encourage sentiment representations of samples from the same class to be pulled closer within each modality. Both image and text embeddings are projected into a shared sentiment embedding space, where the contrastive objectives operate jointly with a standard cross-entropy loss for sentiment classification. This step ensures intra-modal consistency, minimizing discrepancies such as IID and TTD. In the \textbf{second step}, we conduct a conditional judgment on the alignment quality between modalities. Specifically, we compute the similarity matrix between paired image and text sentiment embeddings. If the diagonal consistency score falls below a predefined threshold, we activate an inter-modal supervised contrastive loss $\mathcal{L}_{cl_{m}}$, which further pulls image-text pairs of the same sentiment class closer together, thus reducing ITD and achieving global sentiment-level alignment.

By decoupling intra- and inter-modal sentiment alignment into a two-step training process, SPP-SCL ensures a well-structured sentiment embedding space where \textbf{intra-image, intra-text, and image-text sentiment distances are harmonized}, resulting in more robust and discriminative fused representations.

\subsection{Multi-modal Feature Extraction}
Given an image-text pair sample from an ITSA dataset, denoted as $S = \{I, T\}$, where $I$ and $T$ represent the image and text modalities, respectively, let the corresponding sentiment label be $Y = y_i$, indicating that the sample belongs to the $i$-th sentiment class. Our goal is to propose a framework for balancing sentiment consistency, independent of encoder scale. SPP-SCL is backbone-agnostic and can replace other encoders directly. We focus on widely adopted backbones (BERT-base, ResNet-50) to ensure fair comparison. 

\subsubsection{Image Encoder}
We adopt ResNet-50 \cite{he2016deep} as the image encoder. Given an input image $I$, we extract feature maps from the layer preceding the final global average pooling. These features are then passed through a $1 \times 1$ convolutional layer with batch normalization and ReLU activation, followed by an adaptive average pooling layer and a fully connected projection layer with non-linear activation. The output is the final image representation $\boldsymbol{I}_{p} \in \mathbb{R}^{B \times D}$, where $B$ denotes the batch size and $D$ the embedding dimension. The overall encoding process is defined as:
\begin{equation}
    \label{eq5-1}
    \boldsymbol{I}_{p} = \mathcal{M}\left ( \mathcal{P}\left ( c\left ( \textrm{ResNet}(I) \right )  \right )  \right ),
\end{equation}
where $c(\cdot)$ denotes the $1 \times 1$ convolutional layer, $\mathcal{P}(\cdot)$ the adaptive average pooling operation, and $\mathcal{M}(\cdot)$ the fully connected transformation for dimensionality reduction.

\subsubsection{Text Encoder}
We adopt BERT \cite{devlin2018bert} as the text encoder. Given a text input $T$, we follow prior findings \cite{jawahar2019does,sun2019fine} that fine-tuning only the last four transformer layers (layers 9–12) yields optimal performance for text classification tasks. Accordingly, we extract two types of features from these layers: (1) the [CLS] token embeddings, which capture contextual semantics, and (2) the token-level word embeddings from the corresponding hidden states, which preserve fine-grained lexical information.

To effectively combine these two feature types, we propose a \textbf{Hierarchical Attention (HA)} module that fuses contextual and word-level representations into a unified sentiment embedding. Specifically, let $\boldsymbol{T}_{c} \in \mathbb{R}^{B \times 4 \times 1 \times C}$ denote the [CLS] embeddings and $\boldsymbol{T}_{w} \in \mathbb{R}^{B \times 4 \times N \times C}$ denote the word embeddings, where $B$ is the batch size, $N$ the number of words, and $C$ the hidden dimension.

The HA module first applies one-dimensional global average pooling to $\boldsymbol{T}_{c}$, reducing it to $\mathbb{R}^{B \times 4 \times 1 \times 1}$, and then passes it through a fully connected layer followed by a sigmoid activation $\mathcal{S}(\cdot)$ to generate normalized attention weights across layers. These weights are then element-wise multiplied with $\boldsymbol{T}_{w}$ to produce the weighted word representation $\boldsymbol{T}_{w}^{\prime}$. Afterward, we apply summation and channel-wise pooling operations to aggregate token features across layers and time steps. Finally, the fused features are fed into an LSTM \cite{hochreiter1997long} followed by a projection layer $\mathcal{M}(\cdot)$ to obtain the final text representation $\boldsymbol{T}_{p} \in \mathbb{R}^{B \times D}$. The entire process can be formalized as:
\begin{equation}
    \label{eq5-2}
    \begin{aligned}
        \boldsymbol{T}_{c}^{\prime} &= \mathcal{S} \left( \mathcal{M} \left( \mathcal{P}_{g} ( \boldsymbol{T}_{c} ) \right) \right), 
        \boldsymbol{T}_{w}^{\prime} = \boldsymbol{T}_{c}^{\prime} \odot \boldsymbol{T}_{w}, \\
        \boldsymbol{T}_{p} &= \mathcal{M} \left( \textrm{LSTM} \left( \mathcal{P}_{c} \left( \mathcal{U}( \boldsymbol{T}_{w}^{\prime} ) \right) \right) \right),
    \end{aligned}
\end{equation}
where $\mathcal{S}(\cdot)$ denotes the sigmoid activation, $\mathcal{P}_{g}(\cdot)$ is global average pooling, $\mathcal{P}_{c}(\cdot)$ is channel-wise pooling, $\mathcal{U}(\cdot)$ denotes summation over layers, $\odot$ represents element-wise multiplication, and $\textrm{LSTM}(\cdot)$ is the LSTM network.

\subsection{Cross-Modal Fusion Module}
Effective fusion of visual and textual sentiment cues is essential for sentiment classification in multi-modal contexts \cite{tang2022bafn}. To this end, we propose a Cross-Modal Fusion (CMF) module that integrates the visual representation $\boldsymbol{I}_{p}$ and the textual representation $\boldsymbol{T}_{p}$ into a unified, sentiment-aware feature. 
Inspired by DFAF \cite{8953537}, CMF combines two key components: \textbf{intra-modal feature enhancement} via Self-Attention (SA) \cite{vaswani2017attention} and \textbf{inter-modal interaction} via dynamic attention modulation. The intuition is to dynamically adjust intra-modal attention weights based on complementary information from the other modality, enabling fine-grained sentiment-level interaction rather than simple semantic matching.
Specifically, for both $\boldsymbol{I}_{p}$ and $\boldsymbol{T}_{p}$, we construct respective Query, Key, and Value vectors: $\boldsymbol{Q}_{I}, \boldsymbol{K}_{I}, \boldsymbol{V}_{I}$ for images and $\boldsymbol{Q}_{T}, \boldsymbol{K}_{T}, \boldsymbol{V}_{T}$ for text. During self-attention, we inject inter-modal cues by modulating the query and key vectors via element-wise multiplication with the other modality's global representation (e.g., $\boldsymbol{T}_{p}$ modulates $\boldsymbol{Q}_{I}, \boldsymbol{K}_{I}$). This allows the model to dynamically adjust its attention focus based on sentiment-relevant context from the other modality.
The fusion process is defined as:
\begin{equation}
    \label{eq5-3}
    \begin{aligned}
        \boldsymbol{I}_{m} &= \textrm{Softmax}\left( \frac{ \left( \boldsymbol{Q}_{I} \odot \boldsymbol{T}_{p} \right) \otimes \left( \boldsymbol{K}_{I} \odot \boldsymbol{T}_{p}^\intercal \right) }{\sqrt{D}} \right) \otimes \boldsymbol{V}_{I}, \\
        \boldsymbol{T}_{m} &= \textrm{Softmax}\left( \frac{ \left( \boldsymbol{Q}_{T} \odot \boldsymbol{I}_{p} \right) \otimes \left( \boldsymbol{K}_{T} \odot \boldsymbol{I}_{p}^\intercal \right) }{\sqrt{D}} \right) \otimes \boldsymbol{V}_{T}, \\
        \boldsymbol{M}_{p} &= \left[ \boldsymbol{I}_{m}, \boldsymbol{T}_{m}, \boldsymbol{I}_{p}, \boldsymbol{T}_{p} \right],
    \end{aligned}
\end{equation}
where $\boldsymbol{I}_{m}, \boldsymbol{T}_{m} \in \mathbb{R}^{B \times D}$ are the enhanced modality-specific features, $\odot$ denotes element-wise multiplication, $\otimes$ is matrix multiplication, and $[\cdot]$ denotes concatenation. $\textrm{Softmax}(\cdot)$ denotes the standard attention normalization. The resulting fused representation $\boldsymbol{M}_{p}$ is sentiment-aware and jointly informed by both intra- and inter-modal contextual signals. 

\subsection{Semi-Push-Pull Supervised Contrastive Learning}
As previously discussed, robust image-text sentiment analysis (ITSA) requires a well-structured embedding space where image and text representations of the same sentiment class are closely aligned. This entails minimizing the distance and relationship discrepancies across three perspectives: Image-to-Image (IID), Text-to-Text (TTD), and Image-to-Text (ITD). To this end, we propose a \textbf{Semi-Push-Pull Supervised Contrastive Learning} (SPP-SCL) framework that incorporates both intra-modal and inter-modal alignment strategies. 

\subsubsection{Intra-Modal Supervised Contrastive Learning}
To enforce sentiment consistency within each modality, we introduce two supervised contrastive losses: $\mathcal{L}_{cl_{i}}$ for the image branch and $\mathcal{L}_{cl_{t}}$ for the text branch. Following \cite{khosla2020supervised}, we first construct an \textbf{Intra-Mask matrix} $\mathds{1}^{ij} = \mathds{1}\left [y_{i}= y_{j},i\ne j \right ]$ to identify positive pairs based on sentiment labels.

Let $\boldsymbol{I}_{p}$ and $\boldsymbol{T}_{p} \in \mathbb{R}^{B \times D}$ denote the $\ell_2$-normalized image and text representations. The intra-modal similarity matrix is computed using cosine similarity. The supervised contrastive loss for the image branch is defined as:
{\footnotesize
\begin{equation}
    \label{eq5-4}
    \begin{aligned}
        \mathcal{L}_{cl_{i}} &= -\frac{1}{B}\sum_{i=1}^{B} \frac{1}{B} \sum_{j=1}^{B} \mathds{1}^{ij} \boldsymbol{\textrm{log}}\left ( \frac{\mathit{exp}\left ( s\left ( i,j \right )/\tau \right ) }{\sum_{k=2}^{B} \mathit{exp}\left ( s\left ( i,k \right )/\tau   \right )}  \right ) ,
    \end{aligned}
\end{equation}}
where $s\left ( i,j \right )$ is the cosine similarity between samples $i$ and $j$, and $\tau$ is the temperature scaling factor. The diagonal is excluded since self-similarity is trivial. The loss $\mathcal{L}_{cl_{t}}$ for the text branch follows the same formulation:
{\footnotesize
\begin{equation}
    \label{eq5-5}
    \begin{aligned}
        \mathcal{L}_{cl_{t}} &= -\frac{1}{B}\sum_{i=1}^{B} \frac{1}{B} \sum_{j=1}^{B} \mathds{1}^{ij} \boldsymbol{\textrm{log}}\left ( \frac{\mathit{exp}\left ( s\left ( i,j \right )/\tau \right ) }{\sum_{k=2}^{B} \mathit{exp}\left ( s\left ( i,k \right )/\tau   \right )}  \right ) .
    \end{aligned}
\end{equation}}
These losses effectively reduce intra-modal sentiment variance and ensure that IID $\simeq$ TTD within the sentiment embedding space.

\subsubsection{Inter-modal Supervised Contrastive Learning}
While intra-modal alignment achieves IID $\simeq$ TTD, it does not guarantee cross-modal consistency (\ie ITD alignment). To address this, we propose an inter-modal supervised contrastive loss $\mathcal{L}_{cl_{m}}$.

We define an \textbf{Inter-Mask matrix} $\mathds{1}^{mn} = \mathds{1}[y_m = y_n, m \ne n]$ across image-text pairs. Let $\boldsymbol{I}_{p'}$ and $\boldsymbol{T}_{p'}$ denote $\ell_2$-normalized features extracted after the first training step. The inter-modal similarity matrix is then computed between all image-text pairs. The inter-modal contrastive loss is defined as:
{\footnotesize
\begin{equation}
    \label{eq5-6}
    \begin{aligned}
        \mathcal{L}_{cl_{m}} =-\frac{1}{B}\sum_{m=1}^{B} \frac{1}{B} \sum_{n=1}^{B} \mathds{1}^{mn} \boldsymbol{\textrm{log}}\left ( \frac{\mathit{exp}\left ( s\left ( m,n \right )/\tau \right ) }{\sum_{t=1}^{B} \mathit{exp}\left ( s\left ( m,t \right )/\tau   \right )}  \right ) .
    \end{aligned}
\end{equation}}
This loss encourages image-text samples to align in sentiment space, thereby enforcing IID $\simeq$ TTD $\simeq$ ITD.

\subsubsection{Training Strategy: Two-Step Optimization}
To balance efficiency and alignment quality, we adopt a two-step training strategy. In the first step (\textbf{SPP-SCL$_{1}$}), we jointly optimize $\mathcal{L}{cl_{i}}$, $\mathcal{L}{cl_{t}}$, and the cross-entropy loss $\mathcal{L}_{ce}$. After training, we evaluate whether sentiment distances among all three types (IID, TTD, ITD) are sufficiently consistent based on a threshold \textcolor{red}{$\alpha$}. If not, we proceed to the second step (\textbf{SPP-SCL$_{2}$}), where we fine-tune the model using the inter-modal contrastive loss $\mathcal{L}_{cl_{m}}$ to close any remaining cross-modal gaps. The full procedure is outlined in Algorithm~\ref {algo}.

\subsection{Loss Function}
The overall training loss integrates four components:
\begin{equation}
    \label{eq5-7}
    \begin{aligned}
        \mathcal{L} = \lambda \mathcal{L}_{ce}\left (\boldsymbol{M}_{p},Y \right ) + 
        \mathcal{L}_{cl_{i}} \left (\boldsymbol{I}_{p},Y \right )+\\
        \mathcal{L}_{cl_{t}} \left (\boldsymbol{T}_{p},Y \right )+\mathcal{L}_{cl_{m}} \left (\boldsymbol{T}_{p^{\prime } },\boldsymbol{I}_{p^{\prime } },Y \right ),
    \end{aligned}
\end{equation}
where $\mathcal{L}_{ce}$ represents the cross-entropy loss function, and $\lambda$ is a hyperparameter used to balance the different losses.

\begin{algorithm}[t]
\caption{\PyComment{Two-Step Training Strategy of SPP-SCL (PyTorch Style)}}
\label{algo}
\begin{algorithmic}[1]
\REQUIRE Input image-text pair $(I, T)$, sentiment label $Y$, model $f(\cdot)$, batch size $B$, threshold $\alpha$
\FOR{each batch $(I, T, Y)$ from dataloader}
    \STATE \PyComment{\# Step 1: Intra-modal alignment (SPP-SCL$_1$)}
    \STATE $M_p, I_p, T_p \leftarrow f(I, T)$
    \STATE $\mathcal{L} = \lambda \cdot \mathcal{L}_{ce}(M_p, Y) + \mathcal{L}_{cl_i}(I_p, Y) + \mathcal{L}_{cl_t}(T_p, Y)$
    \STATE Backpropagate and update: $\texttt{optimizer.step()}$

    \STATE \PyComment{\# Step 2: Inter-modal check (SPP-SCL$_2$)}
    \STATE $I_p', T_p' \leftarrow f(I, T)$ \PyComment{\# extract embeddings again}
    \STATE Compute similarity matrix: $\boldsymbol{S} = \texttt{sim}(I_p', T_p')$
    \STATE Compute average diagonal: $d = \texttt{mean}(\texttt{diag}(\boldsymbol{S}))$
    \STATE Count: $c = \texttt{sum}(\boldsymbol{S} < d)$
    \STATE $c_\text{mask} = \texttt{sum}(\texttt{InterMask})$
    
    \IF{$c < \alpha \cdot c_\text{mask}$}
        \STATE $\mathcal{L}_{cl_m} = \mathcal{L}_{cl_m}(I_p', T_p', Y)$
        \STATE Backpropagate and update: \\ $\texttt{optimizer.step()}$
    \ENDIF
\ENDFOR
\end{algorithmic}
\end{algorithm}

\section{Experiments}
\label{S3}
\subsection{Datasets}
We evaluate on three widely used ITSA benchmarks: MVSA-S, MVSA-M \cite{niu2016sentiment}, and the sarcasm-oriented HFM dataset \cite{cai2019multi,liang2022multi}. Details are listed in \tabref{tab5-1}.
\begin{table}[h]
    \renewcommand{\arraystretch}{1}
    \centering
    \setlength{\tabcolsep}{1.2pt}{
    \scalebox{0.8}{
    \begin{tabular}{c|c|c|c|c}
    \hline
    Datasets    & Numbers  & Train sets  & Validation sets & Test sets \\ \hline
    MVSA-S & 4,511  & 3,611  & 450   & 450   \\
    MVSA-M & 17,024 & 13,624 & 1,700 & 1,700 \\
    HFM & 24,635 & 19,816 & 2,410 & 2,409 \\ \hline
    \end{tabular}}}
    \caption{Statistics of three datasets.}
    \label{tab5-1}
\end{table}
\begin{table}[h]
    \renewcommand{\arraystretch}{1}
    \centering
    \setlength{\tabcolsep}{1.6pt}{
    \scalebox{0.7}{
    \begin{tabular}{c|ccc}
    \hline
    \multirow{2}{*}{Hyperparameters}   & \multicolumn{3}{c}{Dataset}   \\ \cline{2-4} 
& \multicolumn{1}{c|}{MVSA-S} & \multicolumn{1}{c|}{MVSA-M} & HFM   \\ \hline
    $\lambda$ & \multicolumn{1}{c|}{5}      & \multicolumn{1}{c|}{5}      & 10    \\
    Initial learning rate                  & \multicolumn{1}{c|}{1e-4}   & \multicolumn{1}{c|}{1e-4}   & 5e-5  \\
    Learning rate decay/multiple             & \multicolumn{1}{c|}{10/0.5} & \multicolumn{1}{c|}{15/0.5} & 5/0.5 \\
    Weight decay                   & \multicolumn{1}{c|}{1e-6}   & \multicolumn{1}{c|}{1e-6}   & 1e-6  \\
    DropOut                & \multicolumn{1}{c|}{0.2}    & \multicolumn{1}{c|}{0.2}    & 0.2   \\
    Batch size                    & \multicolumn{1}{c|}{64}     & \multicolumn{1}{c|}{128}    & 256   \\
    Epochs                 & \multicolumn{1}{c|}{150}    & \multicolumn{1}{c|}{150}    & 150   \\ \hline
    \end{tabular}}}
    \caption{Statistics for all hyperparameters.}
    \label{tab5-2}
\end{table}
\subsection{Experimental Details} 
Our SPP-SCL model is implemented using PyTorch and HuggingFace Transformers, and trained on an NVIDIA V100 GPU (16GB) \footnote{Codes: \url{https://github.com/TomorrowJW/SPP-SCL}}. We use ResNet-50 \cite{he2016deep} and BERT-base-uncased \cite{devlin2018bert} as frozen image and text encoders. Training is performed using the AdamW optimizer \cite{kingma2014adam} with StepLR scheduling. Key hyperparameters include temperature $\tau=0.07$, threshold $\alpha=2/3$, feature dimension $D=32$, max text length $N=200$, and image size $224 \times 224$. Additional settings are listed in \tabref{tab5-2}. To make a fair comparison, we follow the existing evaluation. For the MVSA-S and MVSA-M datasets, the Accuracy (Acc) and Weighted F1 value are used for evaluation, and for the HFM dataset, the Accuracy (Acc) and Macro F1 are used for evaluation.

\begin{table*}[h]
\renewcommand{\arraystretch}{1}
\centering
\setlength{\tabcolsep}{2pt}{
\scalebox{0.75}{
\begin{tabular}{c|c|c|cc|cc|c|c|cc}
\hline
&  &  
& \multicolumn{2}{c|}{MVSA-S}          
& \multicolumn{2}{c|}{MVSA-M}        
&  & \multicolumn{1}{c|}{}   
& \multicolumn{2}{c}{HFM}                   \\
\multirow{-2}{*}{modal}     
& \multirow{-2}{*}{Model}                  
& \multirow{-2}{*}{Publication}         
& ACC& F1 & ACC& F1& \multirow{-2}{*}{Model}                 
& \multirow{-2}{*}{Publication} 
& ACC& F1 \\ \hline
& CNN & EMNLP2014  & 0.6819 & 0.5590  & 0.6564 & 0.5766 
& CNN & EMNLP2014  & 0.8003 & 0.7532 \\
& Bi-LSTM  & ACL2016 & 0.7012 & 0.6506 & 0.6790 & 0.6790
& Bi-LSTM  & ACL2016 & 0.8190 & 0.7753 \\
\multirow{-3}{*}{Text}         
& BERT  & NAACL2018  & 0.7111  & 0.6970  & 0.6759  & 0.6624 
& BERT  & NAACL2018 & 0.8389  & 0.8326 \\ \hline
& ResNet-50  & CVPR2016   & 0.6467  & 0.6155  & 0.6188  & 0.6098 
& ResNet-50  & CVPR2016   & 0.7277  & 0.7138 \\
& OSDA & TMM2021 & 0.6675 & 0.6651 & 0.6662 & 0.6623    
& OSDA & TMM2021    & -  & -  \\
\multirow{-3}{*}{Image}  
& ViT & ICLR2021 & 0.6378 & 0.6226 & 0.6194 & 0.6119 
& ViT & ICLR2021   & 0.7309 & 0.7152 \\ \hline
& MultiSentiNet & CIKM2017   & 0.6984 & 0.6984 & 0.6886 & 0.6811 
& Concat(2) & ACM MM2016   & 0.8103 & 0.7799 \\
& HSAN & ISI2017    & 0.6988 & 0.6690 & 0.6796 & 0.6776 
& Concat(3) & ACM MM2016    & 0.8174 & 0.7874 \\
& Co-MN-Hop6 & SIGIR2018  & 0.7051 & 0.7001 & 0.6892 & 0.6883 
& MMSD & ACL2019  & 0.8344 & 0.8018 \\
& MGNNS & ACL2021    & 0.7377 & 0.7270 & 0.7249 & 0.6934 
& D\&R Net & ACL2020    & 0.8402 & 0.8060 \\
& CLMLF & NAACL2022  & 0.7533 & 0.7346 & 0.7200 & 0.6983 
& CLMLF & NAACL2022  & 0.8543 & 0.8487 \\
& ITIN & TMM2023    & 0.7519 & 0.7497 & 0.7352 & 0.7349 
& HKE & EMNLP2022  & 0.8702 & -  \\
& MULSER & TMM2023    & 0.7560 & 0.7534 & 0.7375 & 0.7371 
& DIP & CVPR2023   & 0.8820 & 0.8767 \\
& CLMLF$^{1}$ & ACL2023    & 0.7378 & 0.7291 & 0.7112 & 0.6863 
& CLMLF$^{1}$ & ACL2023    & 0.8489 & 0.8446 \\
& MVCN & ACL2023    & 0.7606 & 0.7455 & 0.7207 & 0.7001 
& MVCN & ACL2023    & 0.8568 & 0.8523 \\

&CTMWA & TMM2024 
& 0.7196 & 0.7143 & 0.7256 & 0.7157
& MMGCL & TAC2024    & 0.8857 & 0.8870 \\
&MSFN & TOMM2025 
& 0.7898 & 0.7848 & 0.7475 & 0.7262
& ESAM & TMM2025    & 0.9011 & 0.8819 \\
\multirow{-10}{*}{Multi-modal} 
& \cellcolor{newcolor}\textbf{SPP-SCL} & \cellcolor{newcolor}\textbf{2025} 
& \cellcolor{newcolor}\textbf{0.8133} & \cellcolor{newcolor}\textbf{0.8015} & \cellcolor{newcolor}\textbf{0.7871} & \cellcolor{newcolor}\textbf{0.7753} & \cellcolor{newcolor}\textbf{SPP-SCL} & \cellcolor{newcolor}\textbf{2025}          & \cellcolor{newcolor}\textbf{0.9469} & \cellcolor{newcolor}\textbf{0.9450} \\ \hline
\end{tabular}}}
\caption{Quantitative results of the proposed SPP-SCL and these state-of-the-art methods on MVSA-S, MVSA-M, and HFM datasets. The best results are highlighted in \textbf{Bold} in each column.}
\label{tab5-4}
\end{table*}
\subsection{Comparison Methods}
\begin{itemize}
    \item \textbf{Uni-modal baselines}: For text, we include CNN \cite{Kim14}, Bi-LSTM \cite{zhou2016attention}, and BERT \cite{devlin2018bert}; for images, ResNet \cite{he2016deep} and ViT \cite{dosovitskiy2020image}. OSDA \cite{yang2020image}, a multi-view visual classifier, was designed by them.
    \item \textbf{Multi-modal baselines}: On MVSA-S and MVSA-M, we compare with MultiSentiNet \cite{xu2017multisentinet}, HSAN \cite{xu2017analyzing}, Co-MN-Hop6 \cite{xu2018co}, MGNNS \cite{yang2021multimodal}, CLMLF \cite{li2022clmlf}, ITIN \cite{9736584}, MULSER \cite{9920172}, CLMLF$^1$, MVCN \cite{wei2023tackling}, CTMWA \cite{10539274}, and MSFN \cite{zhang2025multimodal}. On HFM, we evaluate against Concat(2), Concat(3) \cite{schifanella2016detecting}, MMSD \cite{cai2019multi}, D\&R Net \cite{xu2020reasoning}, CLMLF \cite{li2022clmlf}, HKE \cite{liu2022towards}, DIP \cite{wen2023dip}, CLMLF$^1$, MVCN \cite{wei2023tackling}, MMGCL \cite{10477507}, and ESAM \cite{yuan2025enhancing}.
\end{itemize}

\subsection{Quantitative Comparison}
The quantitative results in \tabref{tab5-4} show that SPP-SCL consistently outperforms all baseline methods across the three datasets. Overall, multi-modal methods perform better than uni-modal ones, with text-based models outperforming image-based models—likely due to the sparse and noisy nature of visual sentiment cues, which limits the effectiveness of image-only classifiers. 
In a word, SPP-SCL first aligns intra-modal sentiment features independently and then balances inter-modal distributions, leading to more robust sentiment representations and better overall performance.

\subsection{Ablation Study}
\subsubsection{Effectiveness of the two-step strategy}
We evaluate a one-step variant of our model—where all contrastive losses are applied simultaneously—to assess the benefit of the proposed two-step training strategy.
\begin{table}[h]
    \renewcommand{\arraystretch}{1}
    \centering
    \setlength{\tabcolsep}{1.5pt}{
    \scalebox{0.8}{
    \begin{tabular}{c|cc|cc|cc}
    \hline
    \multirow{2}{*}{Method} & \multicolumn{2}{c|}{MVSA-S}  & \multicolumn{2}{c|}{MVSA-M} & \multicolumn{2}{c}{HFM}           \\
                        & ACC             & F1              & ACC              & F1              & ACC             & F1              \\ \hline
    \cellcolor{newcolor}\textbf{SPP-SCL (Two-step)}            & \cellcolor{newcolor}\textbf{0.8133} & \cellcolor{newcolor}\textbf{0.8015} & \cellcolor{newcolor}\textbf{0.7871}  & \cellcolor{newcolor}\textbf{0.7753} & \cellcolor{newcolor}\textbf{0.9469} & \cellcolor{newcolor}\textbf{0.9450} \\ \hline
    SPP-SCL (One-step)          & 0.7956         & 0.7684          & 0.7276           & 0.6929          & 0.8800          & 0.8765          \\ \hline
    \end{tabular}}}
    \caption{Ablation results of two-step strategy.}
    \label{tab5-50}
    \end{table}
\tabref{tab5-50} shows that the one-step strategy performs worse than the two-step strategy, especially on the HFM dataset. This demonstrates that our approach better balances intra- and inter-modal sentiment embedding distances, leading to improved performance.

\subsubsection{Effectiveness of semi-push-pull supervised contrastive learning}
We remove $\mathcal{L}_{cl_{i}},\mathcal{L}_{cl_{t}}$ and $\mathcal{L}_{cl_{m}}$ respectively from the entire SPP-SCL.
\begin{table}[h]
    \renewcommand{\arraystretch}{1}
    \centering
    \setlength{\tabcolsep}{1.5pt}{
    \scalebox{0.8}{
    \begin{tabular}{c|cc|cc|cc}
    \hline
    \multirow{2}{*}{Method} & \multicolumn{2}{c|}{MVSA-S}  & \multicolumn{2}{c|}{MVSA-M} & \multicolumn{2}{c}{HFM}           \\
                        & ACC             & F1              & ACC              & F1              & ACC             & F1              \\ \hline
    \cellcolor{newcolor}\textbf{SPP-SCL}             & \cellcolor{newcolor}\textbf{0.8133} & \cellcolor{newcolor}\textbf{0.8015} & \cellcolor{newcolor}\textbf{0.7871}  & \cellcolor{newcolor}\textbf{0.7753} & \cellcolor{newcolor}\textbf{0.9469} & \cellcolor{newcolor}\textbf{0.9450} \\ \hline
    w/o $\mathcal{L}_{cl_{m}}$       & 0.7289          & 0.7040          & 0.7688           & 0.7500          & 0.7812          & 0.7753          \\
    w/o $\mathcal{L}_{cl_{i}},\mathcal{L}_{cl_{t}}$       & 0.7244          & 0.7170          & 0.7459           & 0.7411          & 0.9427          & 0.9409          \\ \hline
    \end{tabular}}}
    \caption{$\mathcal{L}_{cl_{i}},\mathcal{L}_{cl_{t}}$ and $\mathcal{L}_{cl_{m}}$ ablation results. ``w/o $\mathcal{L}_{cl_{i}},\mathcal{L}_{cl_{t}}$" indicates the removal of $\mathcal{L}_{cl_{i}},\mathcal{L}_{cl_{t}}$, while ``w/o $\mathcal{L}_{cl_{m}}$" indicates the removal of $\mathcal{L}_{cl_{m}}$.}
    \label{tab5-5}
    \end{table}
\tabref{tab5-5} shows that removing either component degrades SPP-SCL’s performance, confirming their importance. On MVSA-S and MVSA-M, intra-modal losses are more critical than inter-modal losses. However, on HFM, the model still performs well without intra-modal loss (F1=0.9409), indicating a larger gap between text and visual features and highlighting the need for $\mathcal{L}_{cl_{m}}$ to balance them.

\begin{figure}[h]
    \centering
    \subfloat[MVSA-S SPP-SCL]{
        \centering
        \includegraphics[scale=0.13]{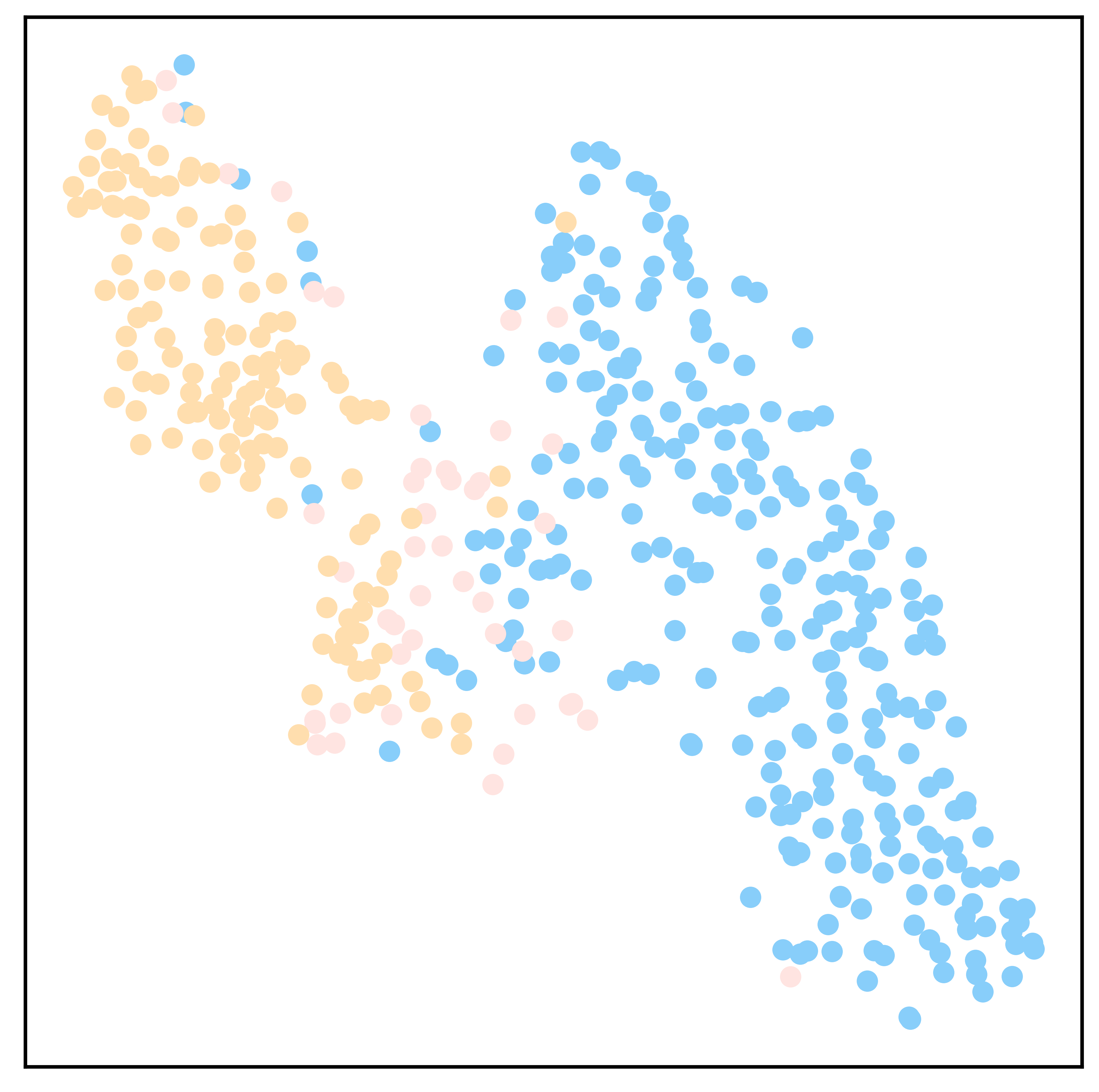}
    }
    \subfloat[MVSA-S without $\mathcal{L}_{cl_{m}}$ ]{
        \centering
        \includegraphics[scale=0.13]{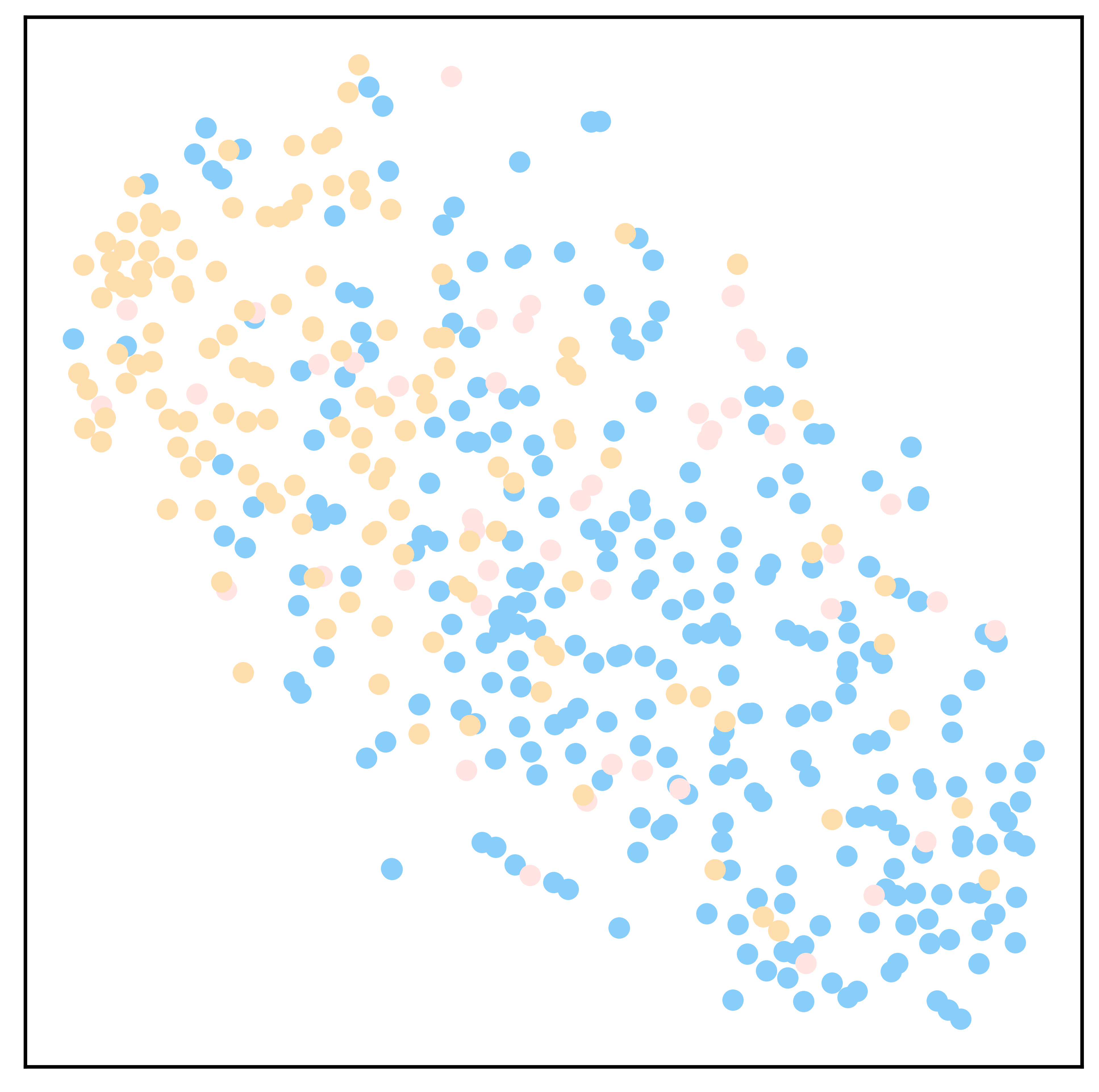}
    }
    \subfloat[MVSA-S without $\mathcal{L}_{cl_{i}},\mathcal{L}_{cl_{t}}$]{
        \centering
        \includegraphics[scale=0.13]{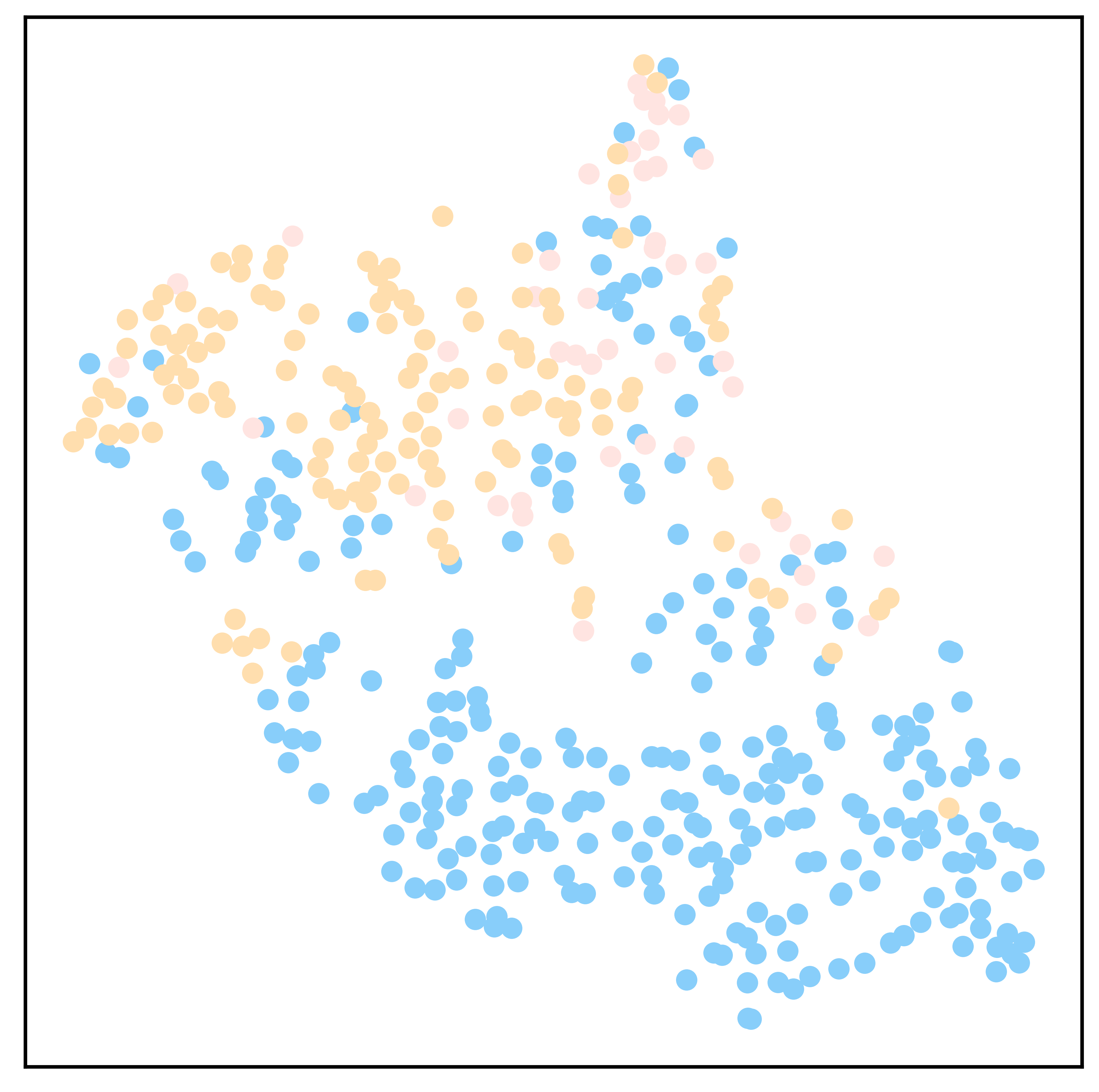}
    }
    \\
    \subfloat[MVSA-M SPP-SCL]{
        \centering
        \includegraphics[scale=0.13]{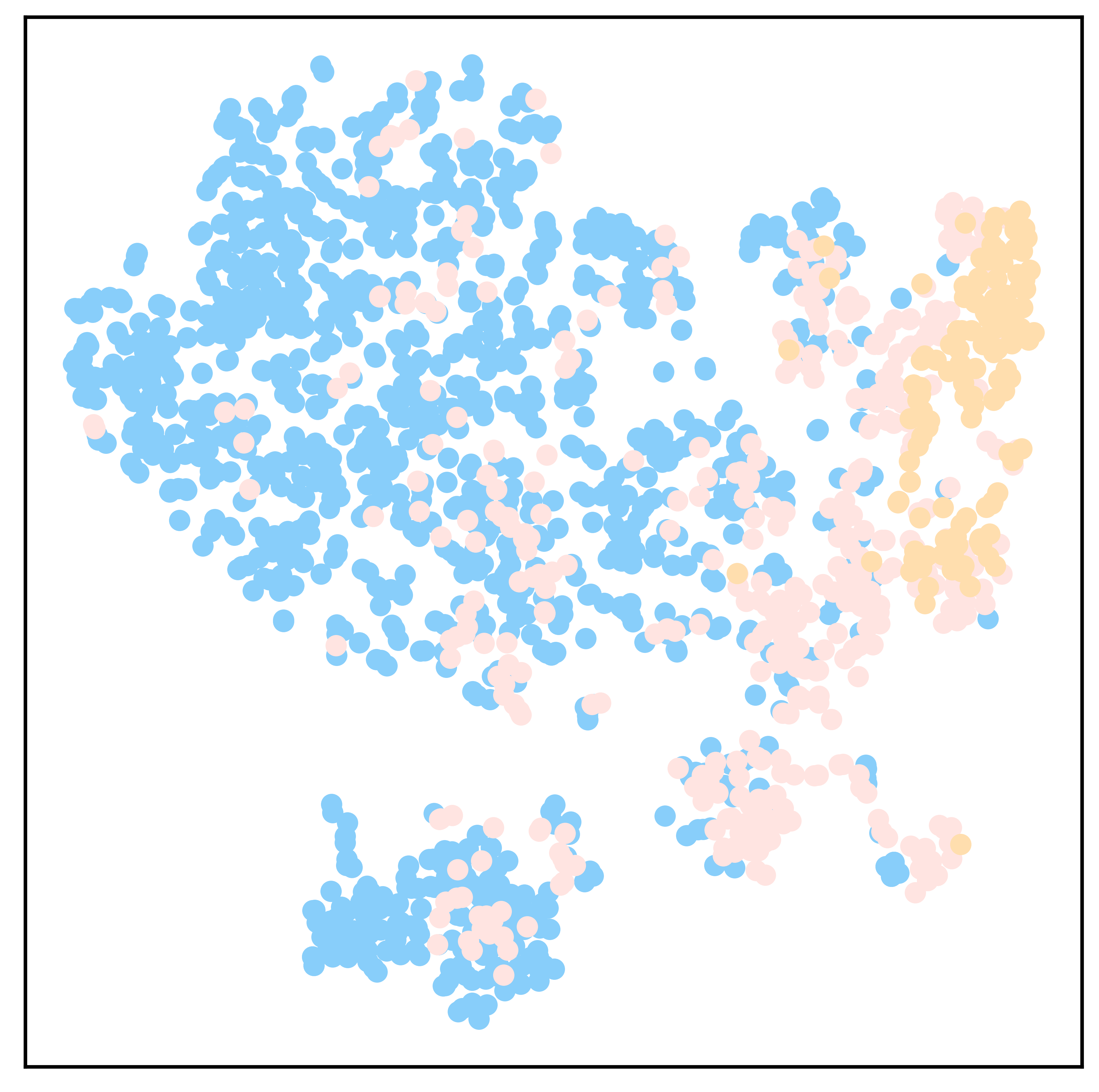}
    }
    \subfloat[MVSA-M without $\mathcal{L}_{cl_{m}}$]{
        \centering
        \includegraphics[scale=0.13]{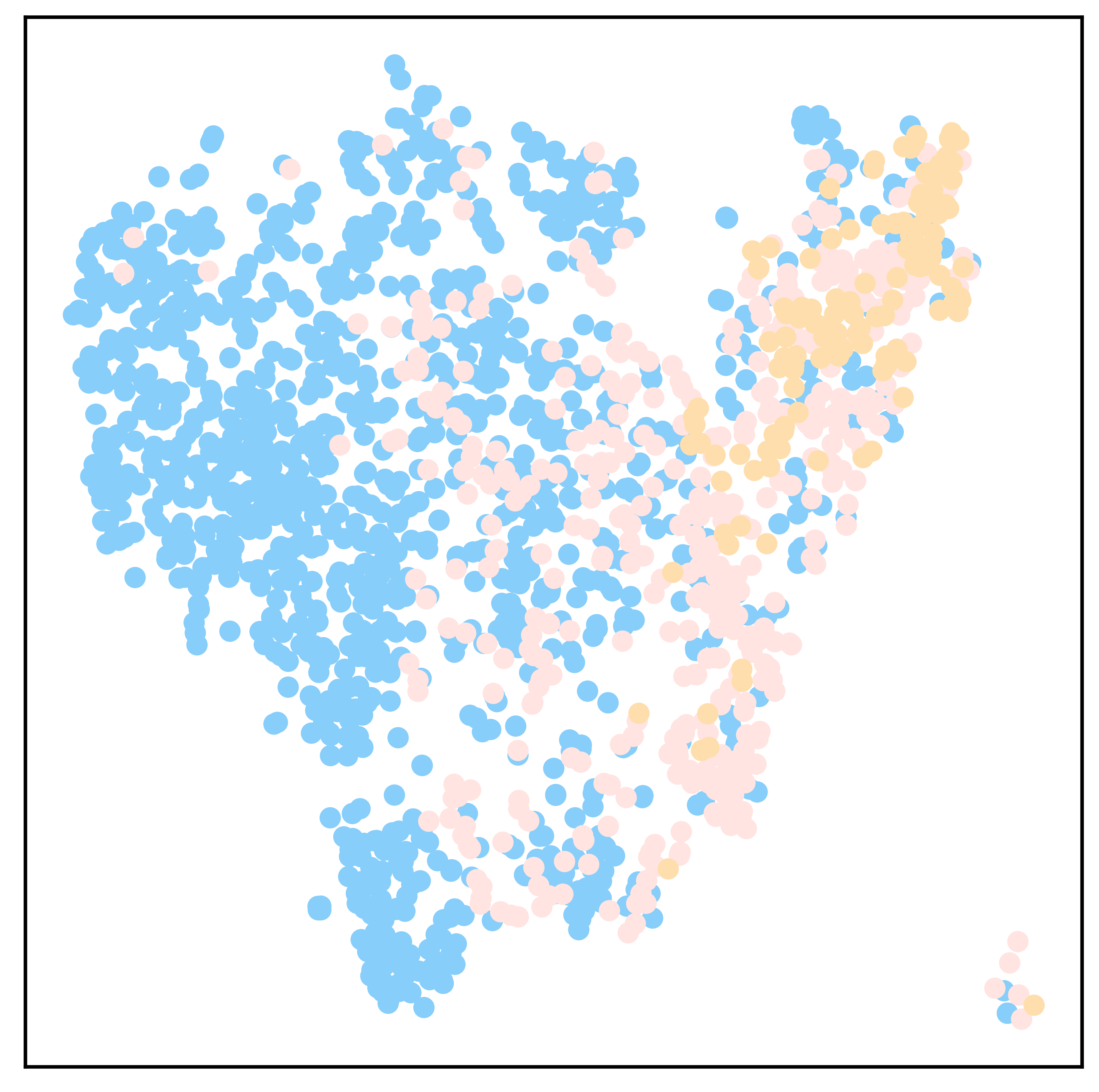}
    }
    \subfloat[MVSA-M without $\mathcal{L}_{cl_{i}},\mathcal{L}_{cl_{t}}$]{
        \centering
        \includegraphics[scale=0.13]{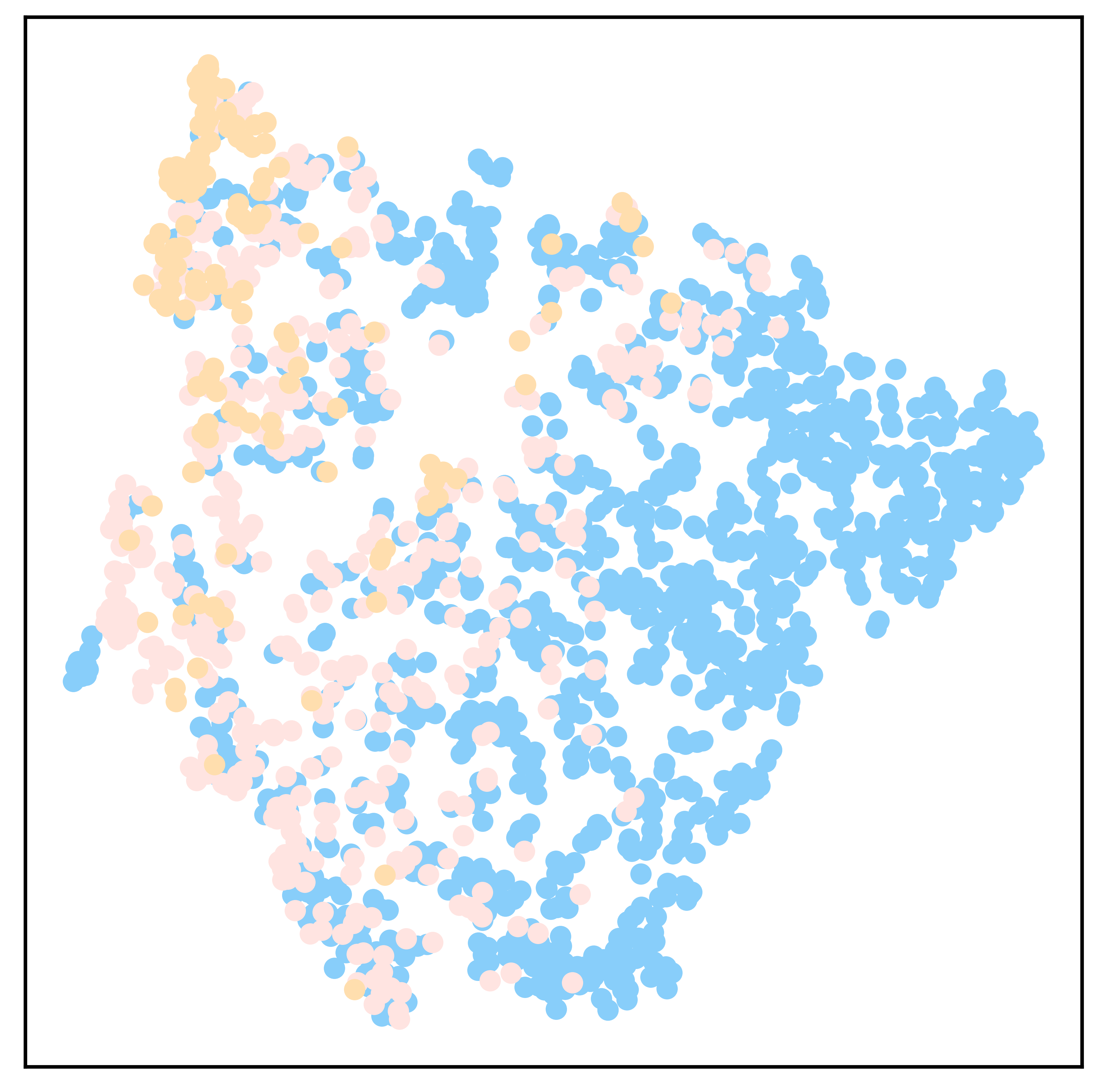}
    }
    \\
    \subfloat[HFM SPP-SCL]{
        \centering
        \includegraphics[scale=0.13]{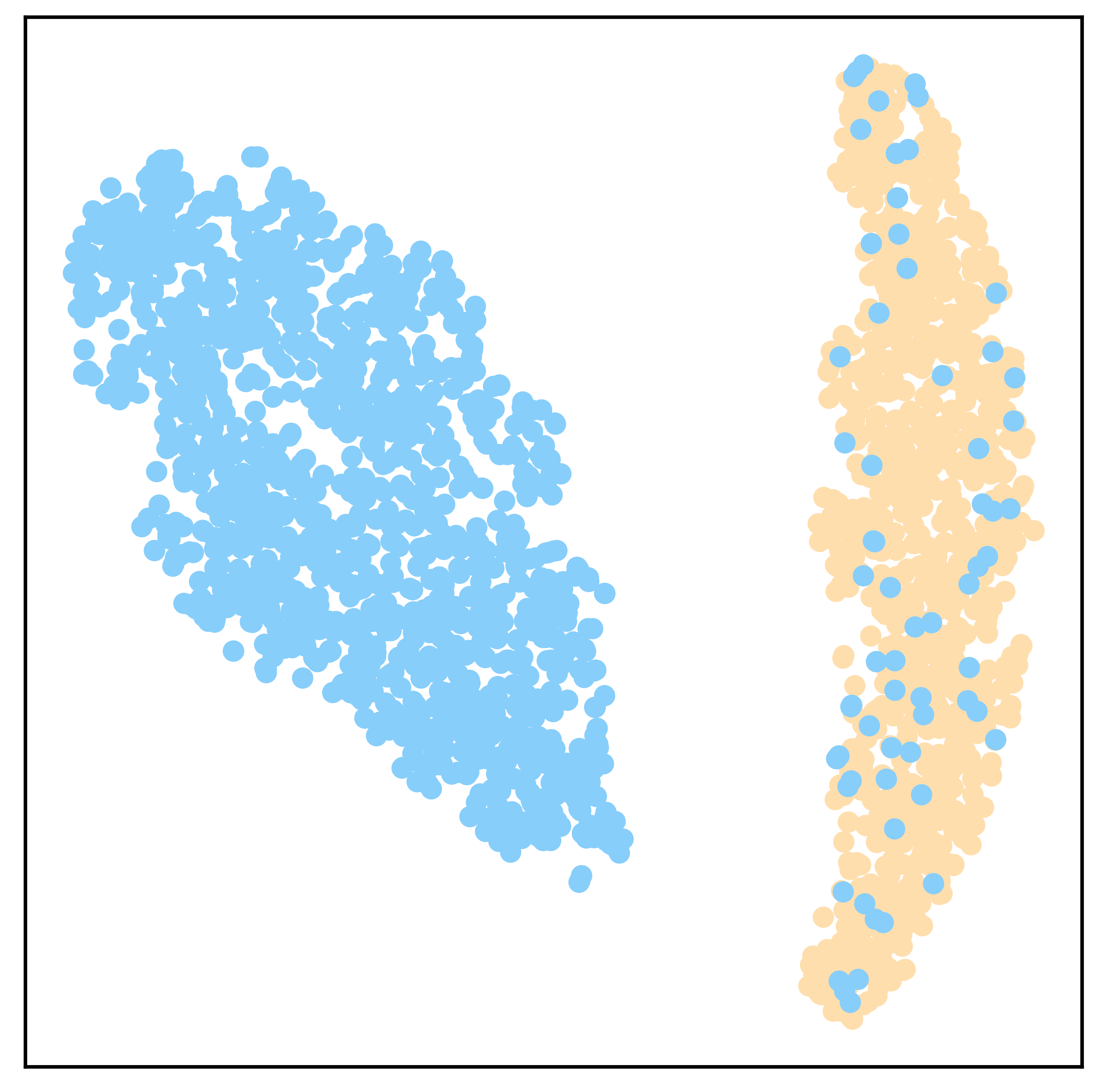}
    }
    \subfloat[HFM without $\mathcal{L}_{cl_{m}}$ ]{
        \centering
        \includegraphics[scale=0.13]{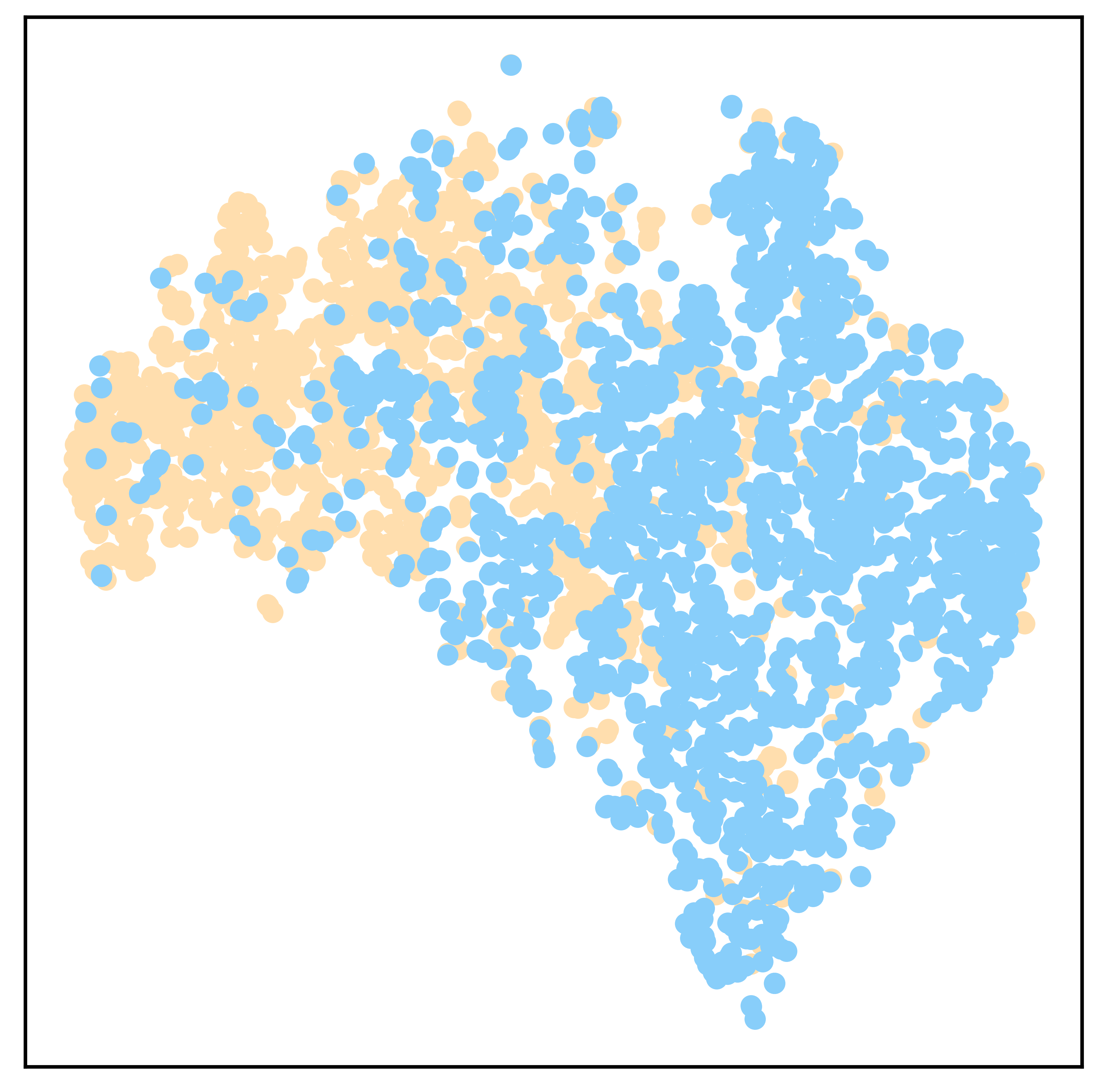}
    }
    \subfloat[HFM without $\mathcal{L}_{cl_{i}},\mathcal{L}_{cl_{t}}$]{
        \centering
        \includegraphics[scale=0.13]{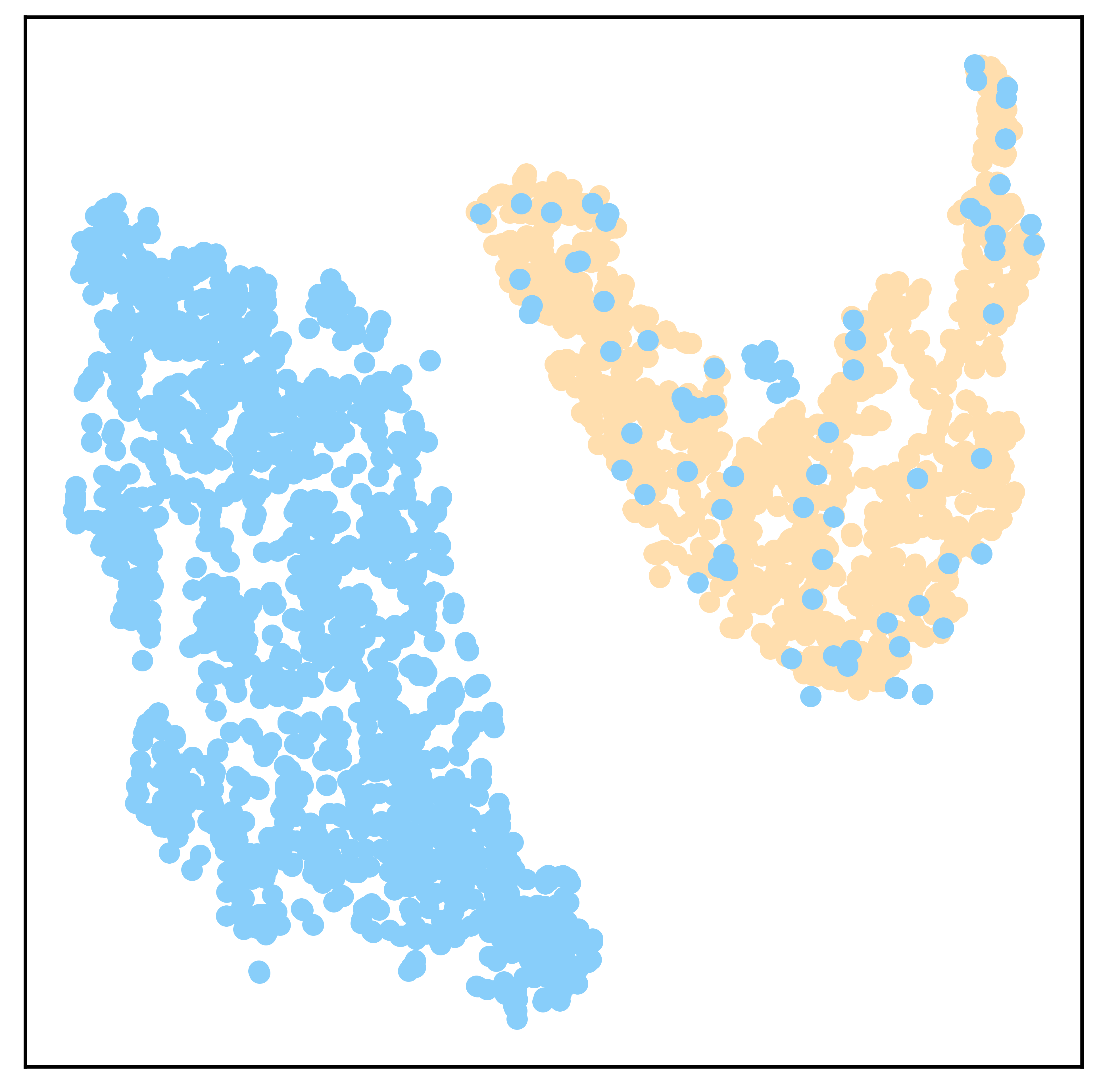}
    }
    \caption{Visualization of the fusion feature distribution on the three datasets.}
    \label{5-6}
\end{figure}

\begin{figure}[h]
    \centering
    \subfloat[Distance distribution of \textcolor{red}{non-sarcasm} category]{
        \centering
        \includegraphics[scale=0.2]{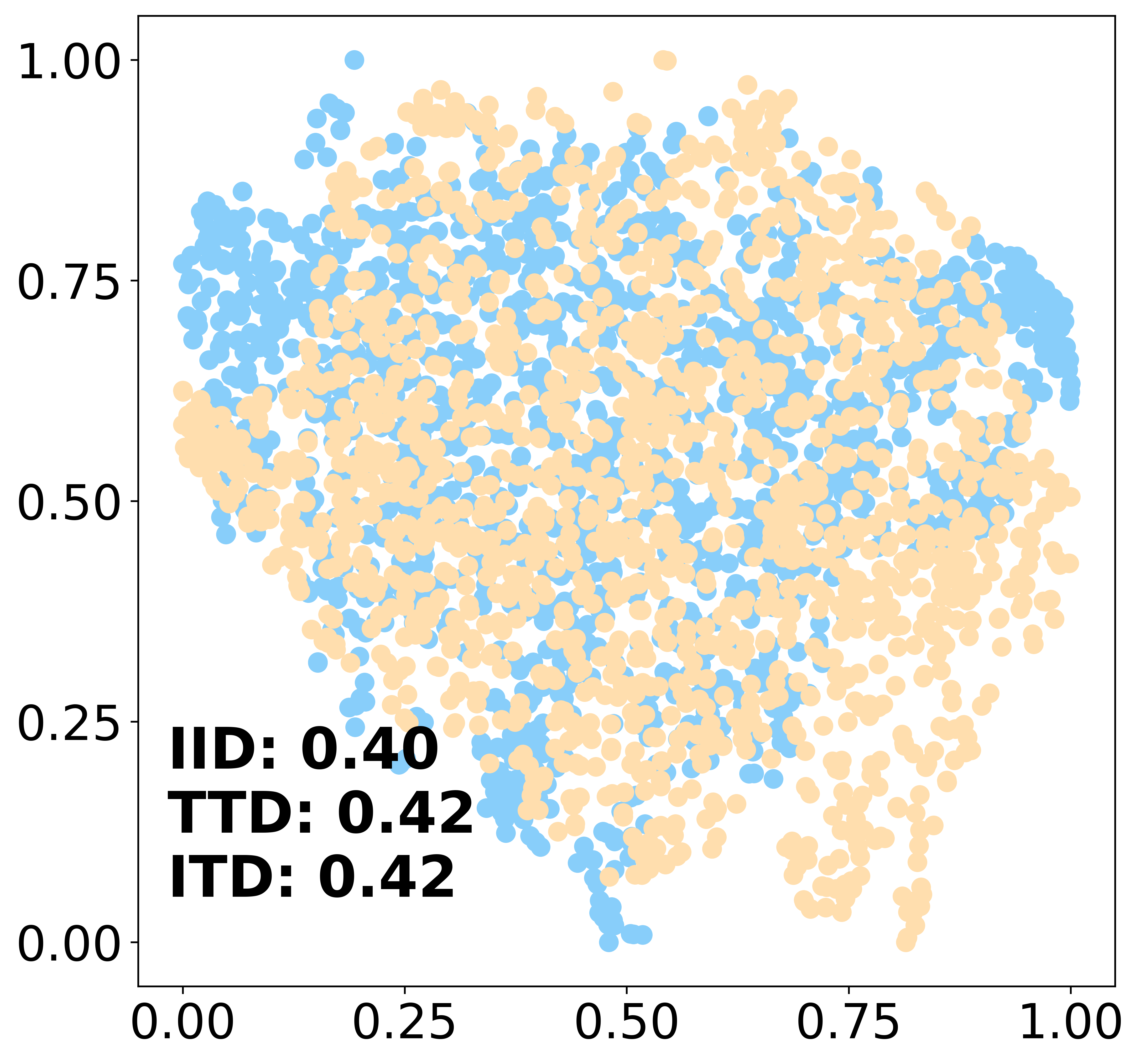}
    }
    \quad
    \subfloat[Distance distribution of \textcolor{red}{sarcasm} category]{
        \centering
        \includegraphics[scale=0.2]{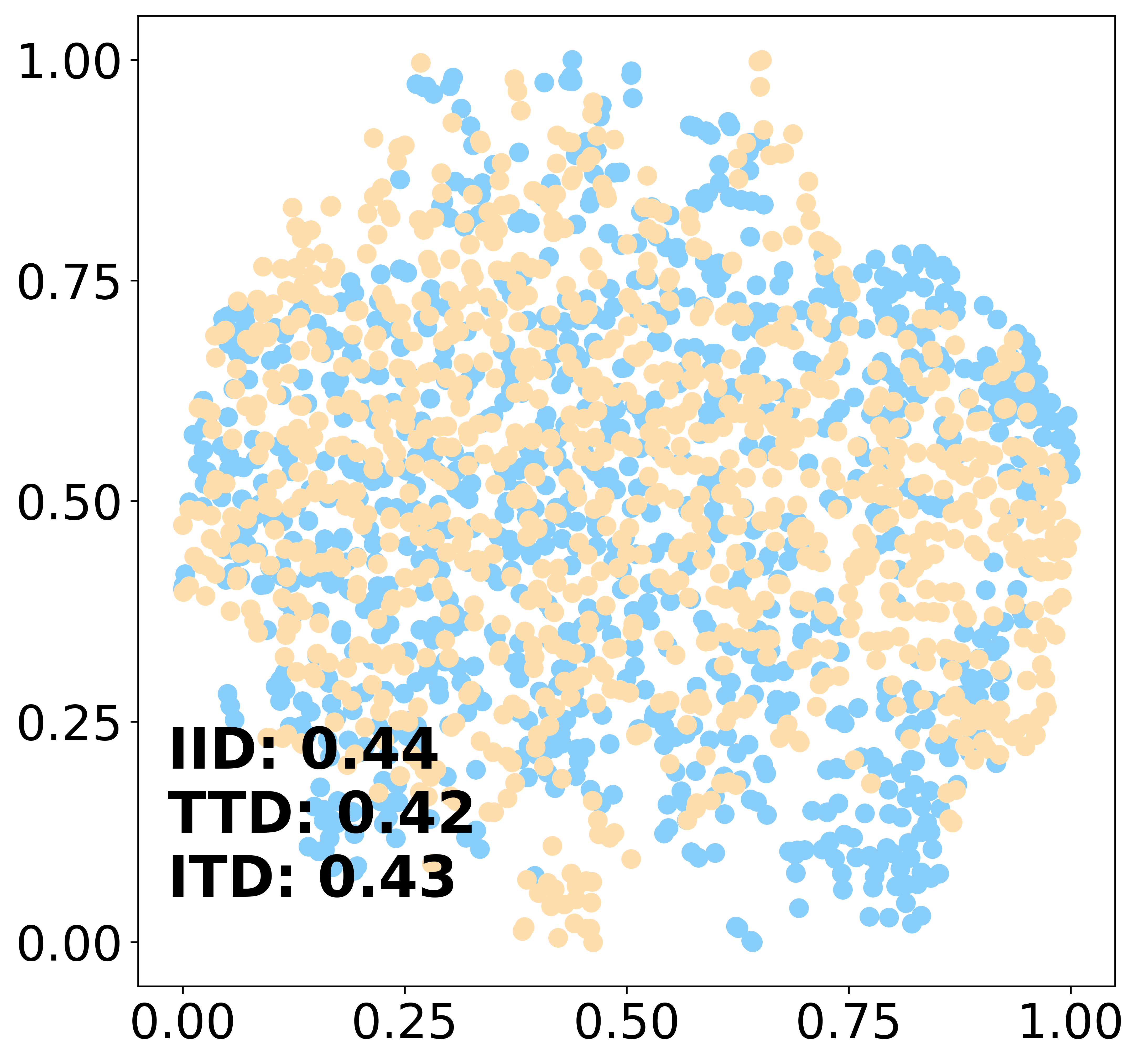}
    }
    \caption{Visualization of the sentiment distance distribution.}
    \label{R-1}
\end{figure}

To further verify the effectiveness of the two components, t-SNE \cite{van2008visualizing} visualizes fusion features before classification in \figref{5-6}. The full SPP-SCL shows well-separated clusters with minimal overlap, while removing $\mathcal{L}_{cl_{m}}$ or $\mathcal{L}_{cl_{i}},\mathcal{L}_{cl_{t}}$ results in larger cluster overlaps. This confirms the components’ role in balancing intra- and inter-modal sentiment consistency.
\figref{R-1} visualizes the distance distribution, showing that average distances (IID, TTD, ITD) are nearly equal and same-category samples from both modalities cluster together. This indicates the sentiment alignment already has notable discriminative power before cross-modal fusion.


\subsubsection{Effectiveness of hierarchical attention module}
To validate the HA module, we conduct ablation experiments extracting [CLS] embeddings from the last 1 to 4 BERT layers and word embeddings from hidden layers. Results are shown in \tabref{tab5-6}. The results show performance varies when using features from BERT’s last four layers across datasets. Generally, fusing all four layers in the HA module gives the best results. MVSA-S, MVSA-M, and HFM are most sensitive to fusing the last two or three layers, but still perform well. This suggests dataset differences, possibly related to size—MVSA-M and HFM are larger. Overall, using the last four layers yields the best performance.

\begin{table}[h]
    \renewcommand{\arraystretch}{1}
    \centering
    \setlength{\tabcolsep}{1.5pt}{
    \scalebox{0.8}{
        \begin{tabular}{c|cc|cc|cc}
            \hline
            \multirow{2}{*}{Method} & \multicolumn{2}{c|}{MVSA-S}  & \multicolumn{2}{c|}{MVSA-M} & \multicolumn{2}{c}{HFM}           \\
     & ACC             & F1              & ACC              & F1              & ACC             & F1              \\ \hline
            w: last three layers              & 0.7667          & 0.7565          & 0.7612           & 0.7275          & 0.9402          & 0.9385          \\
            w: last two layers              & \cellcolor{newcolor}\textbf{0.8222} & \cellcolor{newcolor}\textbf{0.8181} & 0.5918           & 0.5847          & 0.9311          & 0.9293          \\
            w: last one layers              & 0.7133          & 0.7025          & 0.7282           & 0.6840          & 0.6617          & 0.6512          \\ \hline
            \cellcolor{newcolor}\textbf{SPP-SCL}             & 0.8133          & 0.8015          & \cellcolor{newcolor}\textbf{0.7871}  & \cellcolor{newcolor}\textbf{0.7753} & \cellcolor{newcolor}\textbf{0.9469} & \cellcolor{newcolor}\textbf{0.9450} \\ \hline
    \end{tabular}}}
    \caption{The ablation experiment results of the HA module. ``w: last three Layers, w: last two Layers and w: last one Layer" respectively indicate the ablation results from the last three layers, last two layers, and last one layer of BERT.}
    \label{tab5-6}
\end{table}

\subsubsection{Effectiveness of cross-modal fusion module}
To validate the CMF module, we remove it from SPP-SCL and replace fusion with simple concatenation of $\boldsymbol{I}_{p}$ and $\boldsymbol{T}_{p}$. Results are in \tabref{tab5-7}.
\begin{table}[h]
    \renewcommand{\arraystretch}{1}
    \centering
    \setlength{\tabcolsep}{1.5pt}{
    \scalebox{0.8}{
    \begin{tabular}{c|cc|cc|c|cc}
        \hline
        \multirow{2}{*}{Method} & \multicolumn{2}{c|}{MVSA-S}  & \multicolumn{2}{c|}{MVSA-M} & \multirow{2}{*}{Method} & \multicolumn{2}{c}{HFM}           \\
 & ACC             & F1              & ACC              & F1              &                     & ACC             & F1              \\ \hline
        ResNet-50           & 0.6467          & 0.6155          & 0.6188           & 0.6098          & ResNet-50           & 0.7277          & 0.7138          \\
        ViT                 & 0.6378          & 0.6226          & 0.6194           & 0.6119          & ViT                 & 0.7309          & 0.7152          \\ \hline
        w/o CMF             & 0.7289          & 0.7482          & 0.7453           & 0.7389          & w/o CMF             & 0.8117          & 0.8260          \\
        \cellcolor{newcolor}\textbf{SPP-SCL}    & \cellcolor{newcolor}\textbf{0.8133} & \cellcolor{newcolor}\textbf{0.8015} & \cellcolor{newcolor}\textbf{0.7871}  & \cellcolor{newcolor}\textbf{0.7753} & \cellcolor{newcolor}\textbf{SPP-SCL}    & \cellcolor{newcolor}\textbf{0.9469} & \cellcolor{newcolor}\textbf{0.9450} \\ \hline
    \end{tabular}}}
    \caption{The ablation experiment results of the CMF module. ``w/o CMF" indicates the removal of the CMF module.}
    \label{tab5-7}
\end{table}
The results show a performance drop without the CMF module, proving its effectiveness. Simple concatenation fails to capture semantic interactions for effective fusion. Despite this, ``w/o CMF" still outperforms image-only models, confirming the advantage of multi-modal methods.
\subsubsection{Sensitivity of hyperparameters}
\begin{figure}[h]
    \centering
    \includegraphics[scale=0.61]{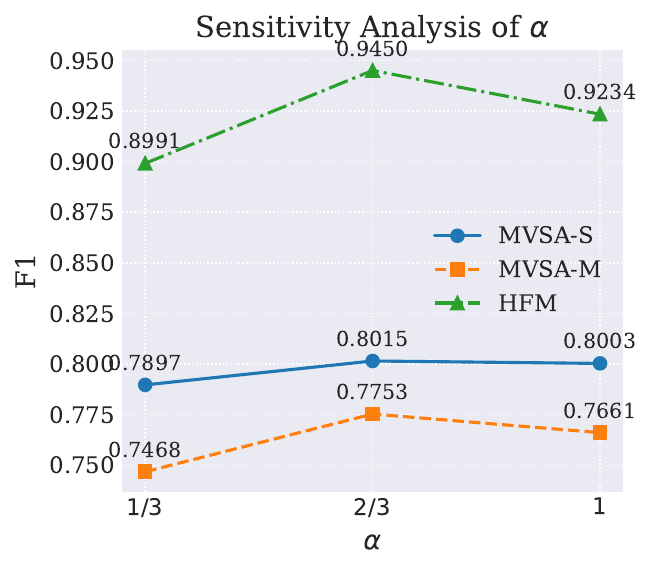}
    \caption{Sensitivity of hyperparameter $\alpha$.}
    \label{ss}
\end{figure}

We observe the best performance at $\alpha = 2/3$, as shown in \figref{ss}. A lower value (\eg $1/3$) triggers inter-modal alignment too early, possibly disrupting already consistent intra-modal structures. In contrast, a higher value (\eg $1$) may skip necessary inter-modal alignment, leading to suboptimal sentiment fusion. Thus, $\alpha = 2/3$ achieves the best trade-off between sensitivity and stability.
\section{Conclusion}
\label{S4}
We propose SPP-SCL, a model that balances visual and textual modalities before fusion for image-text sentiment analysis using a novel two-step supervised contrastive learning strategy, combining intra- and inter-modal learning. Without data augmentation, this approach achieves a balanced sentiment embedding space. We also introduce a hierarchical attention module and a cross-modal fusion module to enhance feature extraction and fusion. Experiments on three datasets validate our method’s effectiveness. Future work will explore semi-supervised image-text sentiment analysis.

\section*{Acknowledgements}
\label{S5}
This work was supported in part by the National Natural Science Foundation of China (No.62541601, No.62306010), Natural Science Foundation of Anhui Province under Grant No. 2408085MF169.

\bibliography{Bibliography-File}

\end{document}